\documentclass[8.5pt,twoside,twocolumn]{article}
\usepackage[super,sort&compress,comma]{natbib} 
\usepackage{mhchem}
\usepackage{times,mathptmx}
\usepackage{sectsty}
\usepackage{balance} 

\usepackage{graphicx} %eps figures can be used instead
\usepackage{lastpage}
\usepackage[format=plain,justification=raggedright,singlelinecheck=false,font=small,labelfont=bf,labelsep=space]{caption} 
\usepackage{fancyhdr}
\begin{document}

\thispagestyle{plain}
\fancypagestyle{plain}{
\fancyhead[L]{\includegraphics[height=8pt]{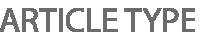}}
\fancyhead[C]{\hspace{-1cm}\includegraphics[height=20pt]{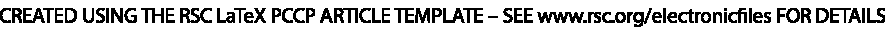}}
\fancyhead[R]{\includegraphics[height=10pt]{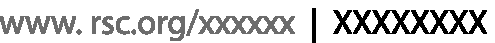}\vspace{-0.2cm}}
\renewcommand{\headrulewidth}{1pt}}
\renewcommand{\thefootnote}{\fnsymbol{footnote}}
\renewcommand\footnoterule{\vspace*{1pt}% 
\hrule width 3.4in height 0.4pt \vspace*{5pt}} 
\setcounter{secnumdepth}{5}

\makeatletter 
\def\subsubsection{\@startsection{subsubsection}{3}{10pt}{-1.25ex plus -1ex minus -.1ex}{0ex plus 0ex}{\normalsize\bf}} 
\def\paragraph{\@startsection{paragraph}{4}{10pt}{-1.25ex plus -1ex minus -.1ex}{0ex plus 0ex}{\normalsize\textit}} 
\renewcommand\@biblabel[1]{#1}            
\renewcommand\@makefntext[1]% 
{\noindent\makebox[0pt][r]{\@thefnmark\,}#1}
\makeatother 
\renewcommand{\figurename}{\small{Fig.}~}
\sectionfont{\large}
\subsectionfont{\normalsize} 

\fancyfoot{}
\fancyfoot[LO,RE]{\vspace{-7pt}\includegraphics[height=9pt]{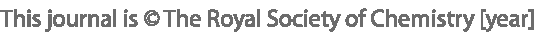}}
\fancyfoot[CO]{\vspace{-7.2pt}\hspace{12.2cm}\includegraphics{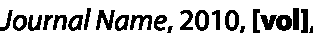}}
\fancyfoot[CE]{\vspace{-7.5pt}\hspace{-13.5cm}\includegraphics{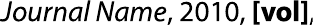}}
\fancyfoot[RO]{\footnotesize{\sffamily{1--\pageref{LastPage} ~\textbar  \hspace{2pt}\thepage}}}
\fancyfoot[LE]{\footnotesize{\sffamily{\thepage~\textbar\hspace{3.45cm} 1--\pageref{LastPage}}}}
\fancyhead{}
\renewcommand{\headrulewidth}{1pt} 
\renewcommand{\footrulewidth}{1pt}
\setlength{\arrayrulewidth}{1pt}
\setlength{\columnsep}{6.5mm}
\setlength\bibsep{1pt}

\twocolumn[
  \begin{@twocolumnfalse}
\noindent\LARGE{\textbf{Domain and Stripe Formation Between Hexagonal and Square Ordered Fillings of Colloidal Particles on Periodic Pinning Substrates }}
\vspace{0.6cm}

\noindent\large{\textbf{Danielle McDermott,\textit{$^{a,b}$} 
Jeff Amelang,\textit{$^{a,c}$}, Lena M. Lopatina,\textit{$^{a}$}, 
Cynthia J. Olson Reichhardt,$^{\ast}$\textit{$^{a}$} and
Charles Reichhardt\textit{$^{a}$}}}\vspace{0.5cm}

\noindent\textit{\small{\textbf{Received Xth XXXXXXXXXX 20XX, Accepted Xth XXXXXXXXX 20XX\newline
First published on the web Xth XXXXXXXXXX 200X}}}

\noindent \textbf{\small{DOI: 10.1039/b000000x}}
\vspace{0.6cm}
%Please do not change this text.

\noindent \normalsize{
Using large scale numerical simulations,
we examine the ordering of colloidal particles on square periodic 
two-dimensional muffin-tin substrates consisting of a flat surface with
localized pinning sites.
We show that when there are four particles per pinning site, 
the particles adopt
a hexagonal ordering, while for five particles per pinning
site, a square ordering appears.
For fillings between four and five particles per pinning site,
we identify a rich variety  
of distinct ordering regimes,  
including disordered grain boundaries, 
crystalline stripe structures, superlattice orderings, 
and disordered patchy arrangements.  
We characterize the different regimes using
Voronoi analysis, energy dispersion, 
and ordering of the domains.  
We show that many of the boundary formation features 
we observe occur for a wide range of other fillings.  
Our results demonstrate that grain boundary tailoring can be 
achieved with muffin-tin periodic pinning substrates. 
}
\vspace{0.5cm}
 \end{@twocolumnfalse}
  ]

\section{Introduction}

\footnotetext{\textit{$^{a}$~Theoretical Division, Los Alamos National Laboratory, Los Alamos, New Mexico 87545, USA. Fax: 1 505 606 0917; Tel: 1 505 665 1134; E-mail: cjrx@lanl.gov}}
\footnotetext{\textit{$^{b}$~Department of Physics, University of Notre Dame, Notre Dame, Indiana 46556, USA. }}
\footnotetext{\textit{$^{c}$~Division of Engineering and Applied Science, California Institute of Technology, Pasadena, California 91125, USA. }}

There is tremendous interest in understanding how to create
different types of self-assembled
colloidal structures. 
It would be particularly useful to identify methods for controlling domain
formation and morphology.
In two-dimensional systems of colloidal particles interacting with 
a repulsive Yukawa potential, the 
lowest energy configuration is a hexagonal lattice \cite{1}, but this
ordering can be modified if the particles interact with a substrate
potential. 
For random substrates, dislocations appear in the colloidal lattice
and destroy the hexagonal ordering \cite{4,9,10}. 
For periodic substrates, the hexagonal ordering may persist, but it
is also possible for
different ordered states as well as
frustrated partially-ordered states to arise due to the competition 
between the
colloidal interactions and the substrate.

For one-dimensional periodic substrates, triangular and smectic type orderings
have been observed \cite{2,3,31,32,33}. 
For two dimensional periodic substrates, the type of ordering 
depends on whether the number of colloidal particles
is greater or less than the number of potential minima.  
In the case of egg-carton substrates, colloidal molecular
crystal states appear whenever the number of 
colloidal particles per potential minima or
the filling factor $f$ is an 
integer with $f=2$ or higher.
Here, the $n$ particles in each potential trap form an    
$n$-mer state with an orientational degree of freedom, and this system
can be mapped to various spin models \cite{4a,5,6,7,11}. 
Another type of two-dimensional 
substrate that has been experimentally realized is 
a muffin tin where the potential minima or pinning sites are
defined as localized trapping sites of 
radius $R_{s}$ surrounded by a flat region 
\cite{8,17,18}.
Experimental realizations of this system 
show that in this case there are two types of colloidal particle
species: the particles that are directly trapped at the pinning sites,
and the remaining particles that are localized on the flat part of the
potential due to their interaction with the directly trapped particles.
These are termed interstitially pinned particles, and they are much more mobile
than the particles in an egg-carton potential \cite{8}.
Muffin tin-type potentials have been studied for 
the ordering and pinning of vortices
in type-II superconductors, where different types of vortex crystalline states 
were shown to occur at integer fillings \cite{13,14,15}.   

At 
noninteger incommensurate fillings, the system
may become disordered. 
For colloidal particles 
in egg-carton potentials at fillings below the $f=1$ matching
filling,
certain crystalline type structures appear at rational fractional filling
ratios
such as $f=1/3$, while at other fillings, the system is either disordered or
else domain wall structures 
form \cite{16}. 
The ordering of repulsive particles for fillings less than one has been
studied for vortices in Josephson junction arrays and in periodic pinning,
where ordered and quasi-ordered patterns arise
\cite{24,25,26,15}. 
Such structures also have many strong similarities to atomic ordering
on periodic substrates 
where the atomic coverage is less than a monolayer \cite{21,19}. 
The incommensurate structures for 
colloidal particles on muffin type structures
at higher filling fractions is not known; 
the additional degrees of freedom in the interstitial 
regions could allow for novel orderings or grain boundary structures
that cannot form at incommensurate fillings on egg-carton potentials.

Here we show that colloidal particles 
at noninteger fillings on muffin tin potentials
exhibit a remarkable
variety of orderings, including a stripe crystalline regime 
dominated by fivefold and sevenfold coordinated particles.
We also find other structures including grain boundaries, superlattices,
and disordered patchy regimes.
We specifically  focus on fillings between four and five particles per trap; 
however, our results     
can be generalized to many other non-integer fillings with more than three
particles per trap, and indicate
that distinct grain boundary morphologies are a 
generic feature of this system. 

\begin{figure}
\centering
\includegraphics[width=0.5\textwidth]{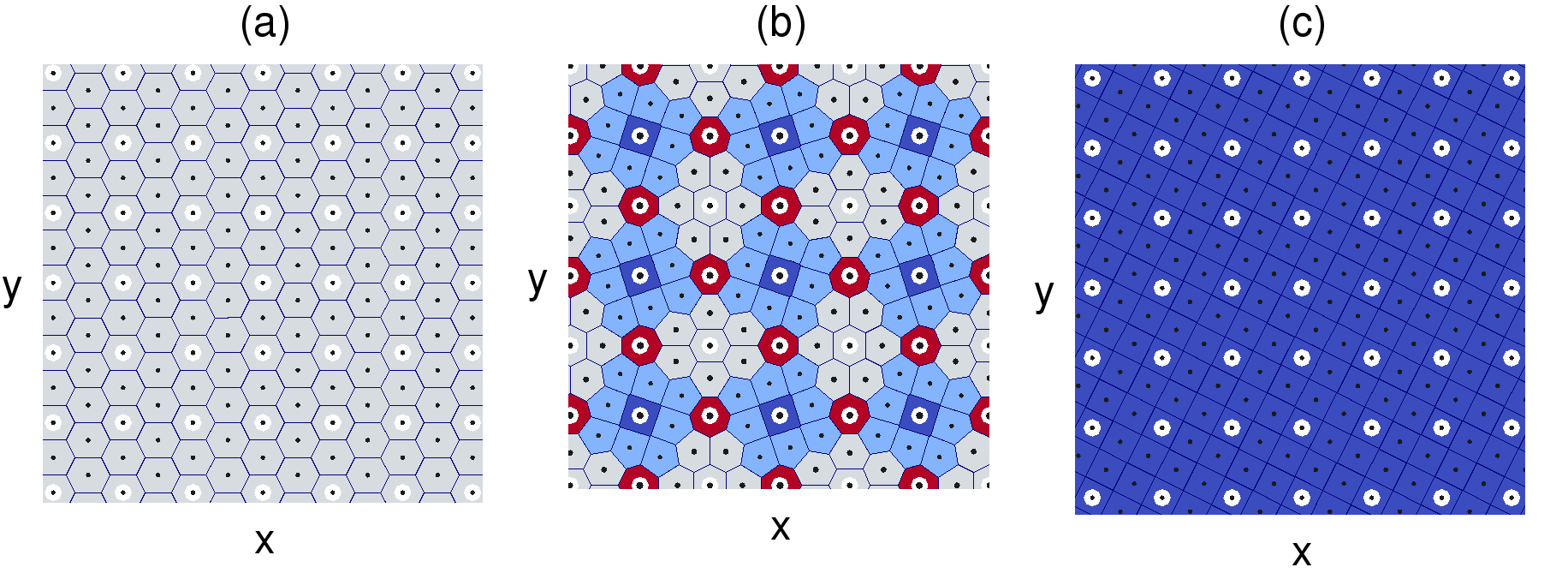}
\includegraphics[width=0.45\textwidth]{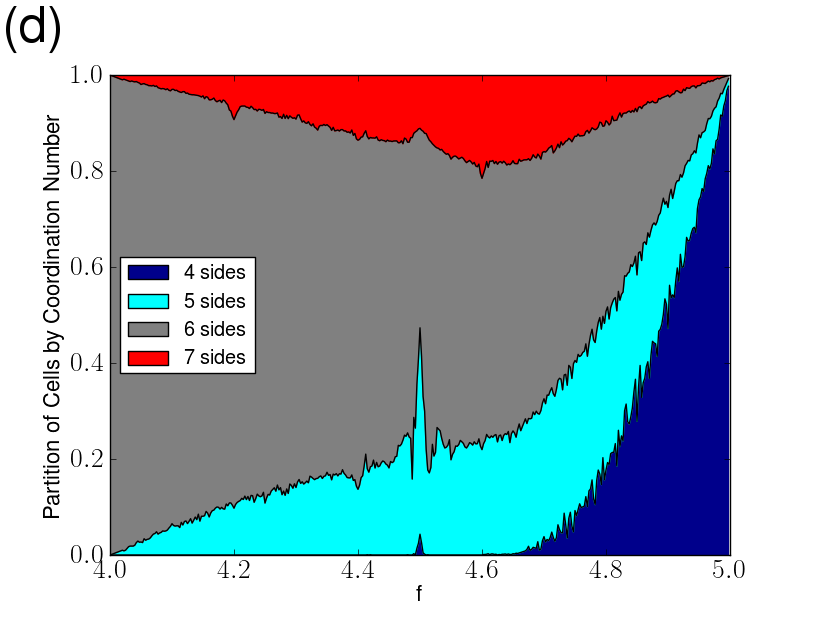}%
\caption{
The Voronoi diagrams of the 
colloidal particle configurations on a square pinning array.
Filled dots: particle locations; open circles: pinning site locations.
Polygon colors indicate the coordination number of the particle at the
center of the polygon: 4 (dark blue), 5 (light blue), 6 (grey), or 7 (red).
(a) At $f=4$, each pinning site captures one particle and
a hexagonal lattice forms.
(b) At $f = 4.5$, the particles form a superlattice 
structure with a periodicity of twice the pinning lattice unit cell.
(c) At $f = 5.0$, a square lattice forms.
(d) A stacked percentage chart of
the fraction of particles with coordination number of 4 (dark blue), 5
(light blue), 6 (grey), and 7 (red) vs filling factor $f$ showing how
the system evolves from a hexagonal structure at $f=4$ to a square
structure at $f=5$.
}
\label{fig:1}
\end{figure}

\section{Simulation and System}

We consider a two-dimensional system 
with periodic boundary conditions in $x$ and $y$ containing 
$N_{c}$ colloidal particles with a density
$\rho = N_{c}/L^2$, where $L$ is the size of the sample in each direction.
The particle-particle interaction is given by
a Yukawa or screened Coulomb potential, and the colloids also
interact with a square array of $N_p$ pinning sites that are each of
radius $R_{p}$. The filling fraction is $f= N_{c}/N_{p}$. 
The 
particle configurations are 
obtained by starting from a high temperature state and performing
simulated annealing to reach
a frozen state.
The motion of particle $i$ during the annealing arises from 
integrating the following overdamped equation of motion:  
\begin{equation}  
\eta \frac{d {\bf R}_{i}}{dt} = 
-\sum_{i\ne j}^{N_{c}}{\bf \nabla}V(R_{ij}) +  {\bf F}^{P}_{i} +   {\bf F}^{T}_{i} 
\end{equation} 
where $\eta$ is the damping coefficient.
The repulsive particle-particle Yukawa interaction potential is  
$V(R_{ij}) =  E_{0}\exp(-\kappa R_{ij})/R_{ij}$,  where
$R_{ij} = |{\bf R}_{i} - {\bf R}_{j}|$, 
${\bf R}_{i(j)}$ is the position of particle $i(j)$,
$E_{0} = Z^{*2}/(4\pi\epsilon\epsilon_{0}a_{0})$, $Z^{*}$ is the effective charge,
$\epsilon$ is the solvent dielectric constant, 
and $1/\kappa$ is the screening length.  
The pinning interaction is modeled as arising from 
non-overlapping parabolic traps with
${\bf F}^{P}_{i} = \sum_{k=1}^{N_p}F_{p}(R_{ik}/R_{p})
\Theta(R_{p}- R_{ik}){\bf \hat{R}}_{ik}$, 
where $R_{ik} = |{\bf R}_{i} - {\bf R}_{k}|$ is the distance between particle 
$i$ and the center of pinning site $k$ located at ${\bf R}_k$,
${\bf {\hat R}}_{ik} = ({\bf R}_{i} - {\bf R}_{k})/R_{ik}$, 
$F_{p}$ is the maximum force of the 
pinning site, and $\Theta$ is the Heaviside step function.   
The effects of thermal fluctuations are modeled by the Langevin term
$F^{T}$ in the form of 
randomly distributed thermal kicks with the properties   
$\langle F^{T}(t)\rangle = 0$ 
and $\langle F^{T}_{i}(t)F(t^{\prime}_{j})\rangle = 2\eta k_{B}\delta_{ij}\delta(t - t^{\prime})$,   
where $k_{B}$ is the Boltzmann constant.
We have tested several annealing rates and work with a rate that is slow enough
that no noticeable difference in the colloidal arrangement appears if
an even slower rate is used.

\begin{figure*}
\centering
\includegraphics[width=0.81\textwidth]{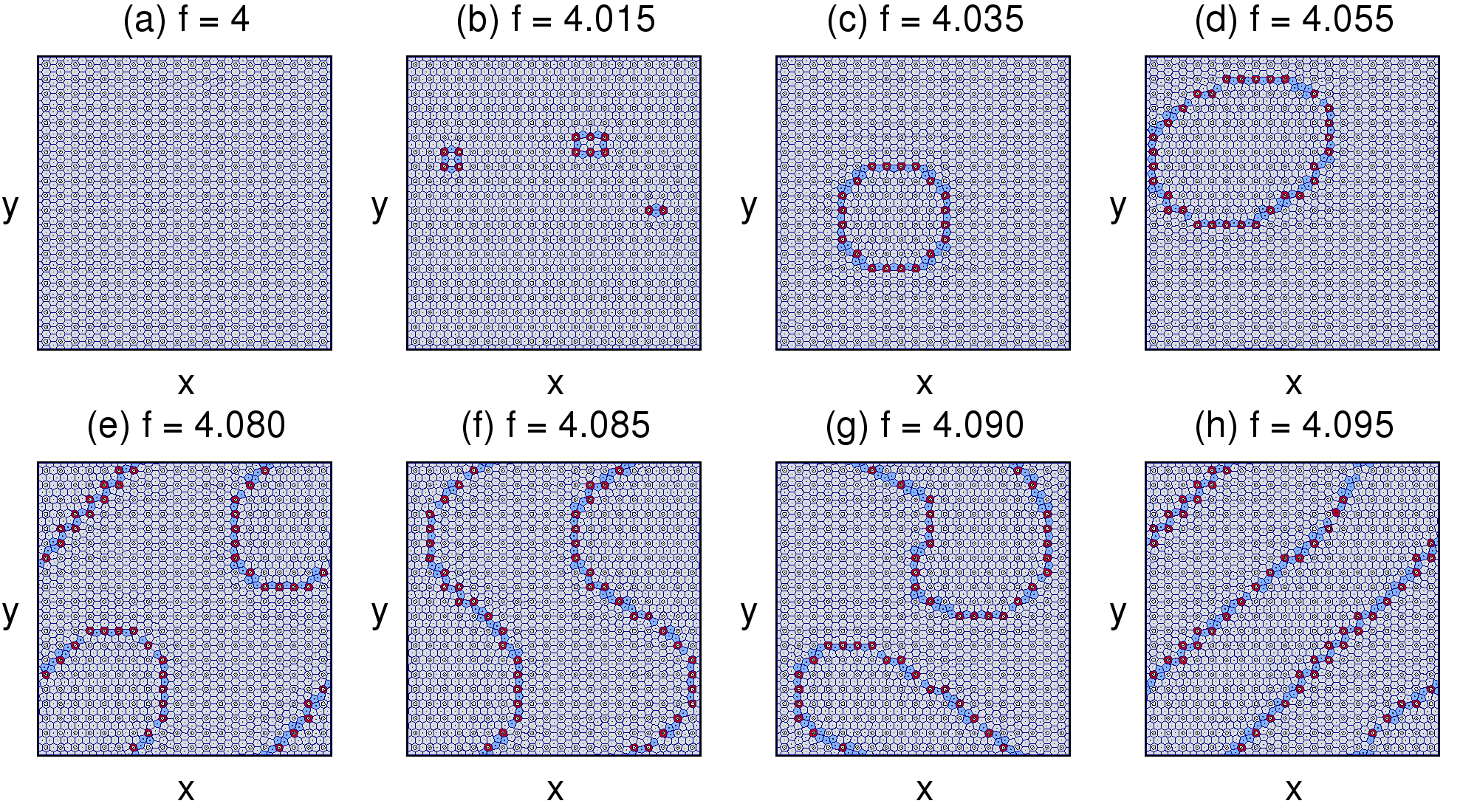}
\caption{
The Voronoi diagrams with the same coloring convention as in Fig.~1
for fillings $f =$ (a) 4.0, (b) 4.015, 
(c) 4.035, (d) 4.055, 
(e) 4.080, (f) 4.085, (g) 4.090, and (h) 4.095.
In this disordered domain wall regime, the particles form grain boundaries
of paired fivefold and sevenfold coordinated (5-7) defects separating different
grain orientations.
}
\label{fig:2}
\end{figure*}

\section{Colloidal Particle Configurations}

We concentrate on systems with $N_p=400$
and $N_c=1600$ to 2000, giving a filling fraction
$4 \leq f \leq 5$. 
At $f = 4$ each pinning site captures one particle while the remaining
particles are located in the interstitial regions.  The result is a 
hexagonal lattice structure in which all of the particles have exactly
six neighbors.
In Fig.~\ref{fig:1}(a) we plot
the Voronoi diagram for the $f = 4$ state indicating the coordination
number of each particle. Here, a hexagonal lattice forms and all particles
are sixfold coordinated, while
at $f = 5.0$, Fig.~\ref{fig:1}(c) shows that the particles
form a square lattice with fourfold coordination. 
At $f = 4.5$ in Fig.~\ref{fig:1}(b), an exotic
isotropic superlattice state appears consisting
of alternating square arrangements of fivefold coordinated 
particles with a fourfold coordinated particle in the center and
groups of six sixfold coordinated particles. 
The unit cell 
of the structure in Fig.~\ref{fig:1}(b)
is twice the size of the pinning lattice unit cell. 

We next address how the system evolves between the 
hexagonal and square orderings as $f$ is varied from $f=4$ to $f=5$.
In Fig.~\ref{fig:1}(d) we plot the fraction of particles with coordination number
4, 5, 6, and 7 for
$4.0 \leq f \leq 5.0$, 
showing that the system begins with hexagonal ordering and has
a peak in the fraction of fivefold and sevenfold 
coordinated particles at $f=4.5$.
For $f > 4.65$ we observe the rapid growth of the fraction of fourfold
coordinated particles with increasing $f$ until the system reaches
a square lattice at $f=5.0$.
We identify four regimes as a function of $f$:
a disordered domain wall regime for 
$4.0 < f < 4.1$, 
a stripe crystal regime for $4.1 \leq f < 4.6$,
a disordered patchy regime for $4.6 \leq f < 5.0$, and
the crystalline states at 
$f = 4$, $4.5$, and $5$. 

In Fig.~\ref{fig:2} we illustrate the domain wall regime for 
$f = 4$, 4.015, 
4.035,
4.055, 4.080, 4.085, 4.090, and $4.095$. 
In this regime, there are no individual defects; instead,
pairs of fivefold and sevenfold coordinated (5-7) defects assemble into
grain boundaries separating
two different orientations of the hexagonal ordering. 
The grain boundaries 
grow in size until they span the entire system, as shown in Fig.~\ref{fig:2}(h)
for $f = 4.095$. 
If we repeat the simulations using different 
initial randomizations of the temperature fluctuations, we obtain
grain boundaries of the same general shape; however, the exact locations
of the grain boundaries can differ.

\begin{figure*}
\centering
\includegraphics[width=0.81\textwidth]{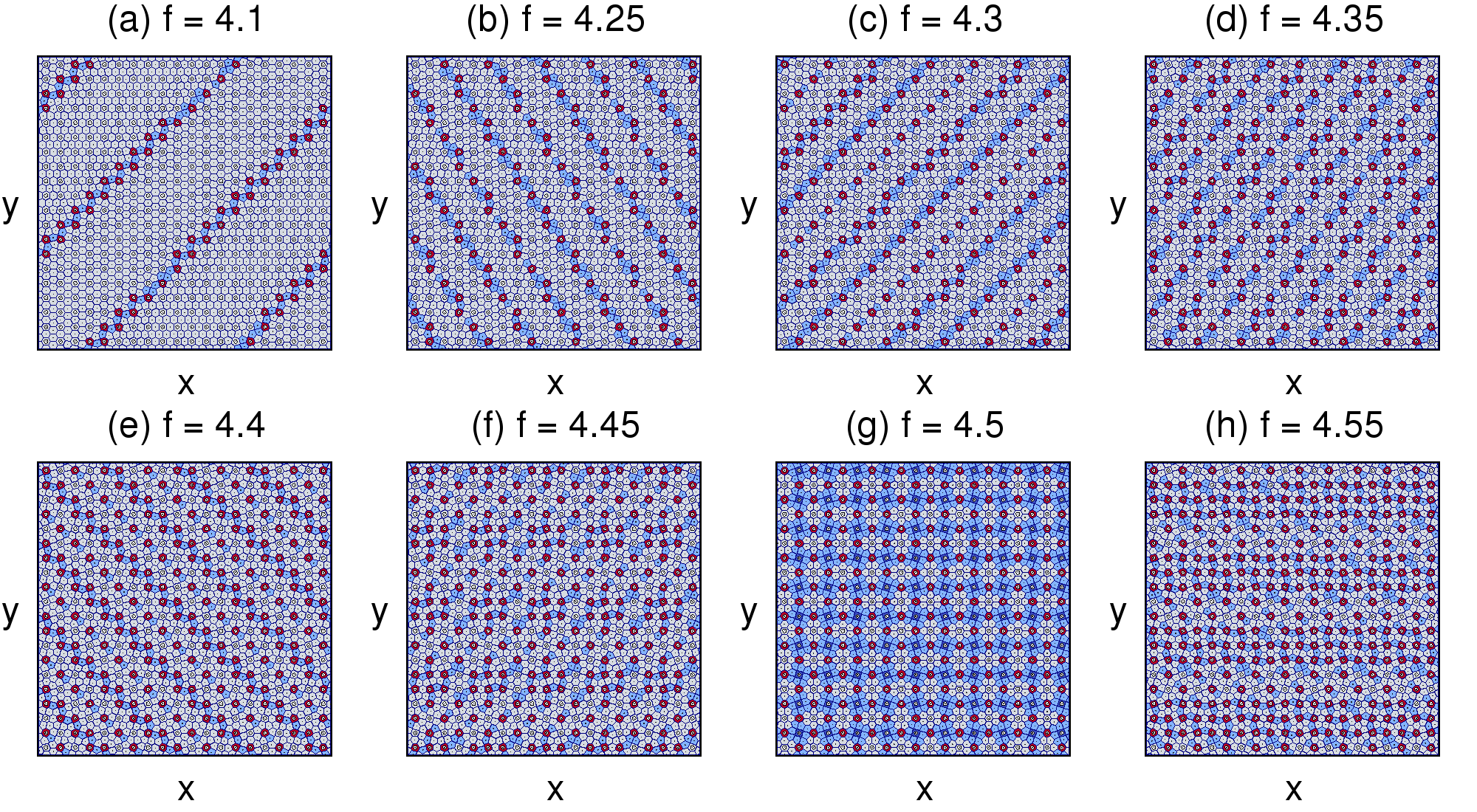}
\caption{
The Voronoi diagrams with the same coloring conventions as in Fig.~\ref{fig:1}
for fillings $f =$ (a) 4.1, 
(b) 4.25, (c) 4.3, (d) 4.35, (e) 4.4, 
(f) 4.45, (g) 4.5, and (h) $4.55$.
Here the system forms stripe crystal states containing ordered lines of 5-7
defects, with the number of stripes increasing with increasing $f$.
The orientation of the stripes with respect to the underlying pinning
lattice varies with filling.
At $f=4.5$ in panel (g), the system forms the isotropic superlattice
structure.
}
\label{fig:3}
\end{figure*}

In Fig.~\ref{fig:3}, we show
particle configurations in the stripe regime
for $f = 4.1$, 
4.25, 4.3, 4.35, 4.4, 4.45, 4.5, and $4.55$. 
Here the system 
forms stripe arrangements of 5-7 defects, where the number of stripes
grows with increasing $f$, and with the isotropic superlattice structure
appearing at $f=4.5$ in Fig.~\ref{fig:3}(g).
As $f$ changes, the orientation of the stripes with respect to the underlying
pinning lattice can change, and the thickness of the stripes can vary.
Some stripe states consist of one-dimensional lines, while in other stripe
states, such as at $f=4.1$ 
in Fig.~\ref{fig:3}(a),
there are zig-zag patterns.
These patterns can even be combined, such as at
$f=4.4$ in Fig.~\ref{fig:3}(e) 
where there are both zig-zag and one-dimensional lines
of stripes.
At $f = 4.45$ and $f=4.55$ in Fig.~\ref{fig:3}(f) and Fig.~\ref{fig:3}(h)
there are two stripe structures that are interspersed. 
We find similar stripe patterns at these fillings in larger samples.
Since the stripe directions are degenerate, it is possible
that domains of different stripe orientation could form in the sample,
as found for the ordering of repulsive particles such as 
superconducting vortices
on periodic arrays for fillings below $f=1$ \cite{15,16}. 
In our system, since most of the particles are not directly pinned
by the pinning sites, they can move freely through the interstitial regions
during the annealing process and can more readily fall into a global
low energy state without different domains.
In contrast, for $f<1$ all the particles are directly
pinned by the substrate and must hop from one site to another via thermal
activation, producing much stronger kinetic constraints and making it 
more likely that domains of different orientation will be quenched into
the sample.
This result indicates that muffin tin potentials permit
much larger ordered regions than can be obtained with egg-carton potentials.

\begin{figure*}
\centering
\includegraphics[width=0.81\textwidth]{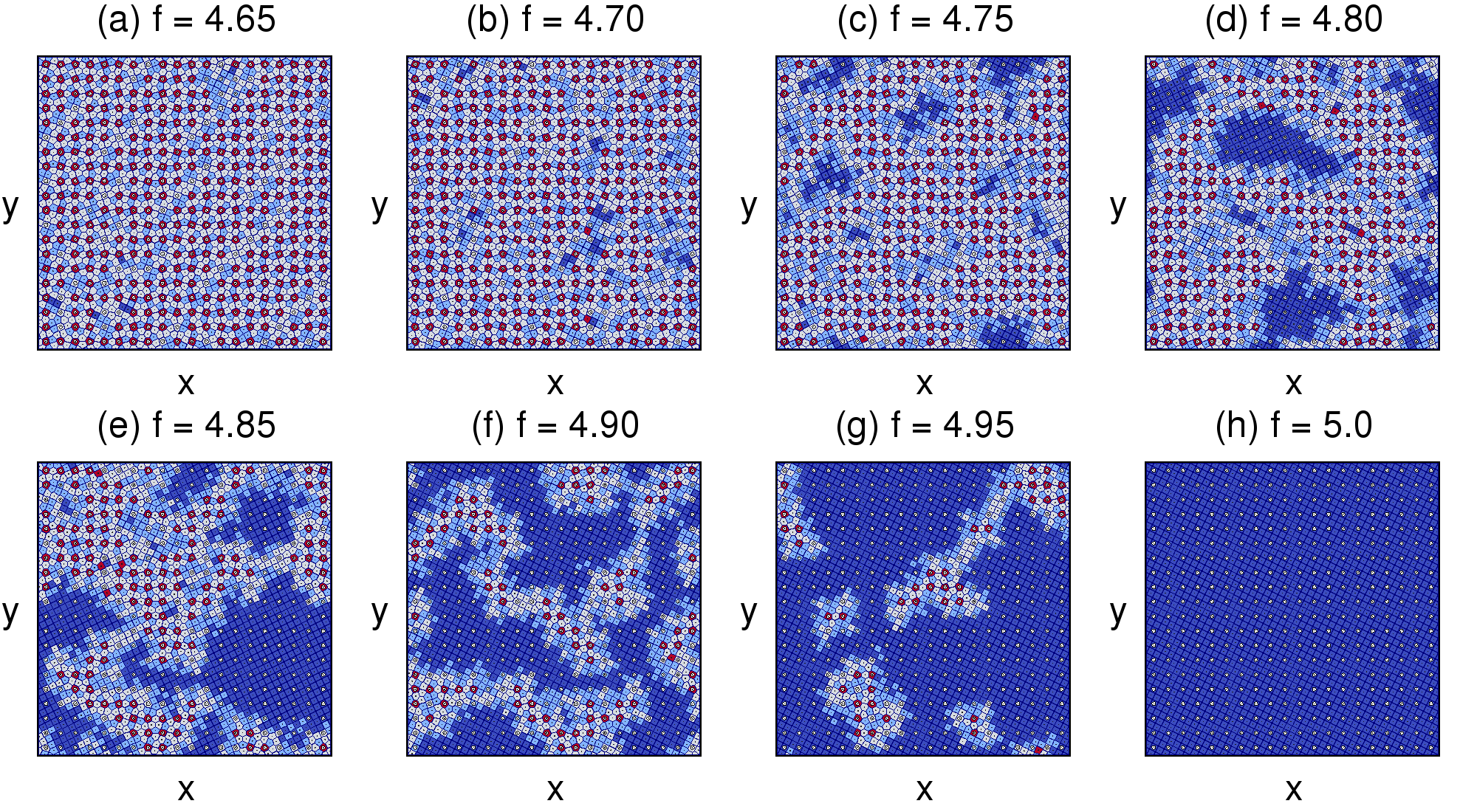}
\caption{ 
The Voronoi diagrams with the same coloring conventions as in Fig.~1
for fillings $f =$ 
(a) 4.65, (b) 4.7, (c) 4.75, (d) 4.8, (e) 4.85, 
(f) 4.9, (g) 4.95, and (h) $5.0$. 
Here the stripe structure is replaced by disordered
patchy regions of square domains that grow in size with increasing $f$.    
}
\label{fig:4}
\end{figure*}

Figure \ref{fig:4} shows the evolution out of the stripe state into the 
disordered patch state at 
$f =$ 
4.65, 4.7, 4.75, 4.8, 4.85, 4.9, 4.95, and $5.0$. 
The stripe state is still present at
$f = 4.6$; 
however, at $f = 4.65$ in Fig.~\ref{fig:4}(a), patches of 
disorder 
begin to emerge, and at $f = 4.70$ in Fig.~\ref{fig:4}(b) 
some fourfold coordinated particles
appear in patches that increase in size with increasing $f$ 
before finally filling
the entire system at $f=5.0$ in Fig.~\ref{fig:4}(h).
The patches do not form regular stripelike patterns as seen for 
$f<4.6$ but are more disordered, implying that the 
low energy state
in this regime must be much
more degenerate than in the stripe forming regime.

\begin{figure}
\centering
\includegraphics[width=0.42\textwidth]{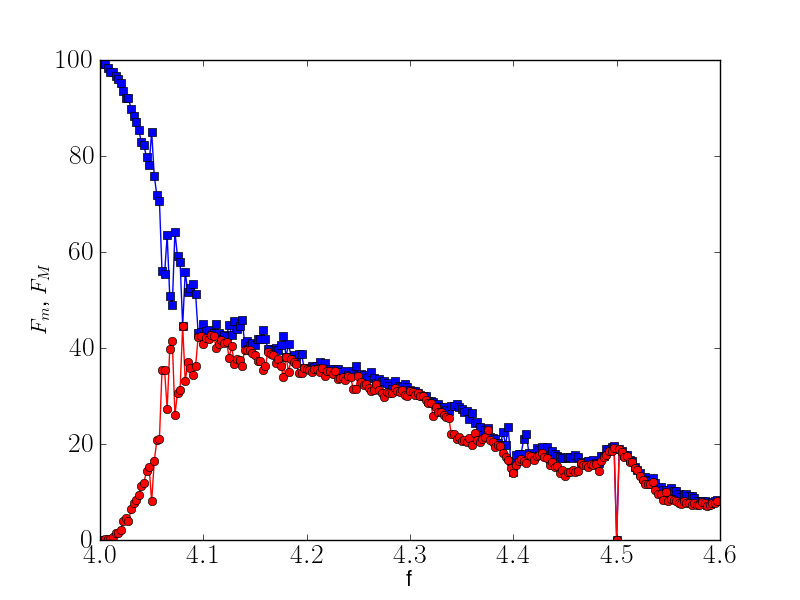}
\includegraphics[width=0.42\textwidth]{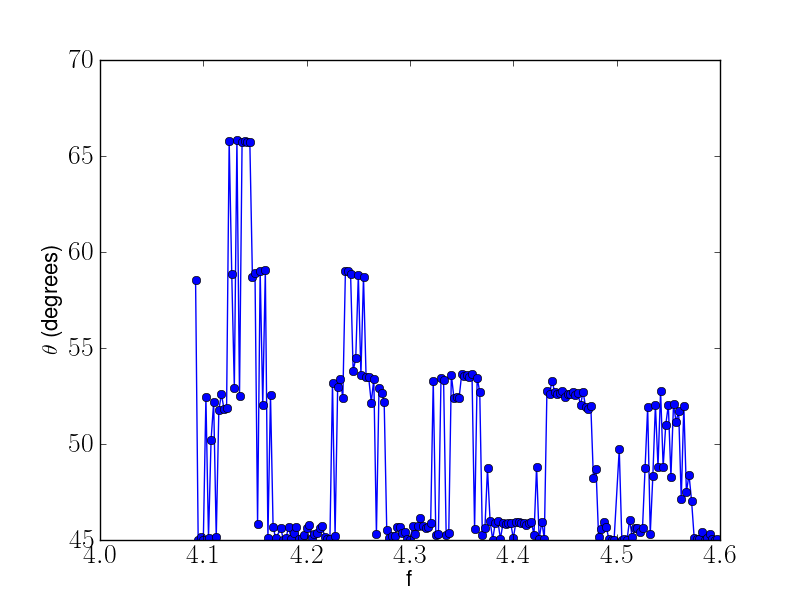}
\caption{
(a) The fraction of particles in minority grains $F_m$ (circles) 
and majority grains $F_M$ (squares) vs $f$.
For $ f < 4.1$ the system is in the domain wall regime. 
When the two curves meet, the system
enters the stripe regime. 
For $f > 4.6$, stripes can no longer be identified and the system
enters the disordered patch regime. 
(b) Distribution of the stripe orientation 
angle $\theta$ with respect to the $x$ axis vs $f$ in the
stripe regime. 
The stripes are predominantly aligned along $\theta=45^\circ$, 
$53^\circ$, $59^\circ$, and $66.8^\circ$, corresponding to symmetry directions of
the underlying square pinning array.
}
\label{fig:5}
\end{figure}

\section{Analysis}
In order to quantify where the different regimes occur, 
we have developed an algorithm to identify the grain boundaries and their
orientation \cite{jeffpaper}.
In the domain wall regime for $4.0 < f < 4.1$,
the sixfold coordinated particles in each domain
can take one of two degenerate orientations. 
We define the majority grain direction to be 
the orientation of the largest domain in the sample, and 
the minority grain direction as the other degenerate orientation.
We then measure the fraction of particles $F_M$ that reside in domains
with the majority orientation and the fraction of particles $F_m$ in domains
with the minority orientation.
This definition is necessary since the majority grain direction can vary
from one disorder realization to another.
Figure \ref{fig:5}(a) shows $F_M$ and $F_m$ versus $f$.
Just above $f = 4$,
there are small patches of minority grains, 
with most of the particles in one large majority grain.
As $f$ increases, $F_M$ and $F_m$ approach each other until
meeting near $f = 4.1$ where the stripe regime begins. 
The stripe regime consists of 
crystalline states that have approximately equal quantities of 
grains of each orientation. 
The overall fraction of sixfold coordinated particles drops with
increasing $f$
as more stripe walls appear. 
Our grain boundary identification algorithm breaks down for $f > 4.6$ 
when the disordered patches appear, and also 
at $f = 4.5$ in the isotropic superlattice. 

In Fig.~\ref{fig:5}(b) we plot the distribution of the 
angle $\theta$ between the stripe orientation 
and the $x$ axis for the fillings
$4.1 \leq f < 4.6$ where the stripes can be identified.
There are four predominant angles, $\theta=45^\circ$,
$53^\circ$, $59^\circ$, and $66.8^\circ$.
These angles can be written as
$\theta=\tan^{-1}(p/q)$ with integer $p$ and $q$ for
$p/q = 1$, 4/3, 5/3, and $7/3$, indicating that the stripes are aligning
with symmetry directions of the underlying square pinning array.
The other observed values of $\theta$ can also be matched to higher
order rational ratios of $p/q$.
The alignment of particle structures 
with the symmetry directions of an underlying pinning lattice 
has also been observed for
colloidal particle ordering on quasicrystalline arrays \cite{27} 
as well as for driven colloidal particles
moving over periodic and quasicrystalline   
arrays \cite{30,29,28}.  
Not all angles corresponding to possible rational values of $p/q$ 
can be realized due to the fact that the domain walls are composed
of 5-7 defects and therefore have a finite thickness.

\begin{figure}
\centering
\includegraphics[width=0.5\textwidth]{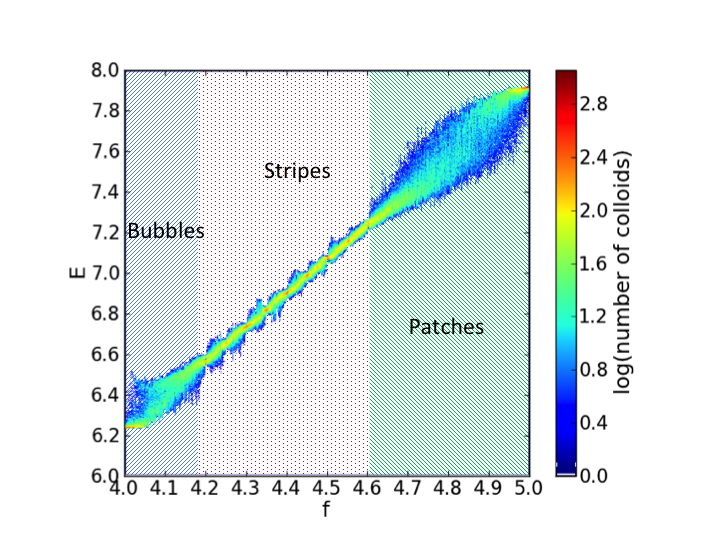}
\caption{ 
The energy distribution per particle $P(E)$ vs $f$ plotted using a color
scheme where red indicates high values of $P(E)$ and blue indicates low
values of $P(E)$. 
In both the grain boundary regime and the disordered patch regime, 
$P(E)$ is broad, 
while in the crystalline stripe
regime, $P(E)$ is much narrower. 
}
\label{fig:6}
\end{figure}

For each filling, we can determine the particle-particle
interaction energy of each particle
$E_i=\sum_{i\neq j}^{N_c}V(R_{ij})$ and then construct the distribution $P(E)$
of the energy of all particles in the system.
In Fig.~\ref{fig:6} we plot the resulting $P(E)$ versus $f$.
For $f < 4.1$, $P(E)$ has a maximum 
just above $E = 6.2$,   
corresponding to the per-particle energy associated with 
the hexagonal lattice at $f = 4.0$. 
As $f$ increases, the peak in $P(E)$ remains close to $E=6.2$ with some
weight in $P(E)$ shifting to higher energies due to the increasing fraction
of grain boundaries in the sample.
The highest energy particles are located along the grain boundaries.
Once the system enters the stripe regime for $4.1 \leq f < 4.6$, $P(E)$
becomes much narrower due to the strong crystalline ordering of the stripe
phases, while the peak energy increases linearly with increasing
$E$.
In the disordered patch regime, $P(E)$ broadens again,
indicating that the particles form disordered states 
with many different particle-particle interaction energies. 
Just below and at $f = 5.0$, $P(E)$ becomes very narrow again
as the square lattice forms.  

We can compare the stripe ordering 
to the pattern formation found on 
periodic egg-carton potentials such as vortex ordering
in superconducting wire networks or colloidal particles on 
optical trap arrays for fillings $f < 1.0$ \cite{15,16,24,25,26}.
In these network systems, the $f=1/3$ filling for a square array 
produces a stripelike pattern of particles.
This is contrast to our system, where the stripes are not individual particles 
but domains of non-sixfold coordinated particles.
In the network systems, the ordering from 
$0 \leq f \leq 1/2$ has a duality with $1/2 \leq f \leq 1$, 
where the configurations for $f=1/2 - x$ and $f=1/2 + x$ 
are the same except that particle locations are
replaced by void locations.
In our system for $4.0 < f <5.0$, a similar duality does not occur.
Additionally, at the higher fillings, the interstitial regions in 
the muffin tin potential
allow particles to take positions that are 
inaccessible in egg-carton potentials.   

\begin{figure}
\centering
\includegraphics[width=0.5\textwidth]{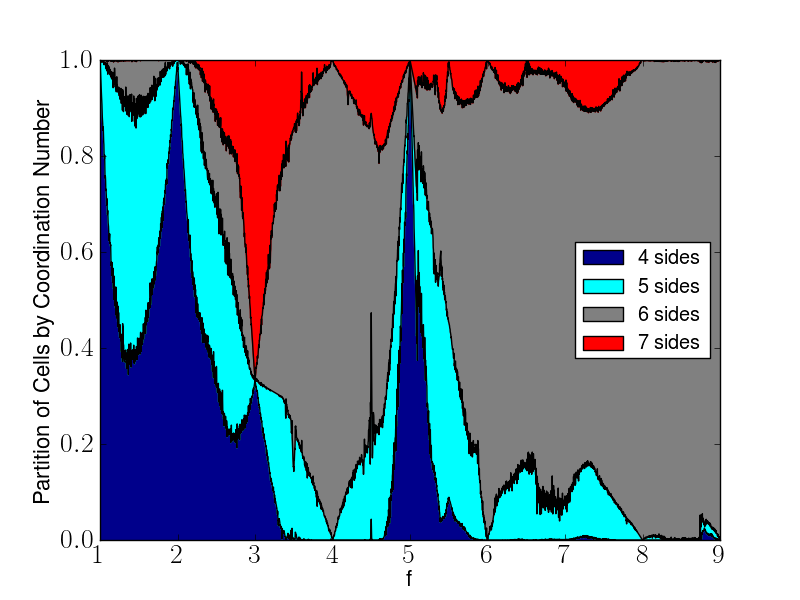}
\caption{A stacked percentage chart of the fraction of particles with
coordination number of 4 (dark blue), 5 (light blue), 6 (grey), and 7 (red)
vs $f$ over the range $1.0 \leq f \leq 9.0$.  Square lattices appear at
$f=1.0$, 2.0, and 5.0, while hexagonal lattices form at $f=4.0$, 6.0,
8.0, and 9.0.
}
\label{fig:7new}
\end{figure}

\begin{figure*}
\centering
\includegraphics[width=0.81\textwidth]{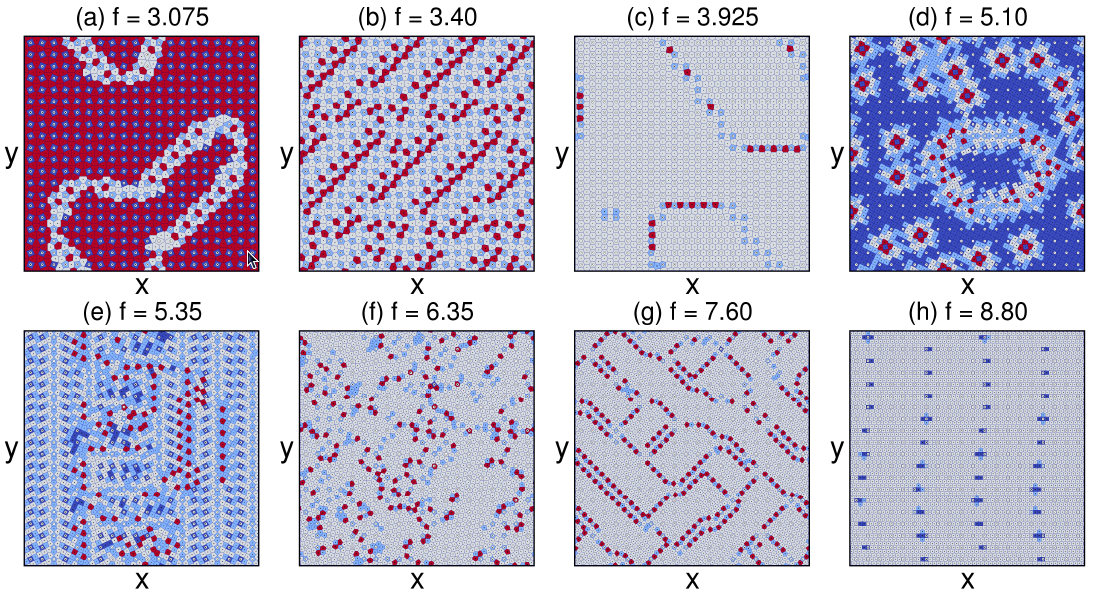}
\caption{The Voronoi diagrams with the same coloring conventions as in
Fig.~\ref{fig:1} for other fillings.
(a) At $f = 3.075$, grain boundaries appear. 
(b) At $f = 3.4$, stripes form.
(c) At $f = 3.925$, grain boundaries again form.
(d) At $f = 5.1$ there is a disordered patch state. 
(e) At $f = 5.35$ a stripe state forms. 
(f) At $f = 6.35$ the system is disordered. 
(g) At $f = 7.6$, labyrinth type grain boundaries form. 
(h) At $f=8.8$, a particle stripe state appears.
}
\label{fig:7}
\end{figure*}

\section{Other fillings}

We have also examined incommensurate fillings from 
$f=1.0$ to $f=9.0$. 
In Fig.~\ref{fig:7new} we show the fraction of particles with coordination
number 4, 5, 6, and 7 versus $f$ over this entire range.
Square lattices occur at $f=1.0$, 2.0, and 5.0, with 100\% of the particles
fourfold coordinated at these fillings, while hexagonal lattices occur
at $f=4.0$, 6.0, 8.0, and 9.0, with 100\% of the particles sixfold
coordinated at these fillings.
At $f=7.0$ the system forms a patchy nonordered pattern, while
at $f=3.0$ the interstitial colloidal particles form a dimer crystal 
arrangement with alternating
dimer orientation.
Figure~\ref{fig:7new} shows considerable fine structure at noninteger
fillings, and certain types of superlattices form for some fillings
as indicated by pronounced peaks and dips at $f=3.5$, 4.5, and 5.5.
In general, we find that domain wall formation 
occurs at most incommensurate fillings; however, the
morphology can exhibit differences depending on the closest integer
filling.
The best examples of stripe crystalline
states appear between $f=4$ and $f=5$.
In the range $f=2$ to $f=3$,
the system forms only disordered
patch states, and no grain boundaries could be identified, 
while between $f=3$ and $f=4$, 
the system exhibits grain boundary and stripe regimes
but no disordered patchy regime.
The ordering at $f= 3$ is not triangular or square but is instead a
superlattice state.
Figure~\ref{fig:7}(a) shows the grain boundary state at $f =  3.075$, 
Fig.~\ref{fig:7}(b) shows the stripe state at  $f = 3.4$, 
and Fig.~\ref{fig:7}(c) shows that at $f=3.925$
a grain boundary state appears that is 
similar to that found at $ f = 4.05$. 
For $ 5 < f < 6$, the system exhibits disordered patch states
of the type illustrated in Fig.~\ref{fig:7}(d) for $f =5.1$, 
as well as stripe states where
the stripes are aligned only along $\theta=0^\circ$ or $\theta=90^\circ$,
as shown in Fig.~\ref{fig:7}(e) for $f = 5.35$. 
For $6 < f < 7$, the system does not form stripes but is instead
disordered, as shown in Fig.~\ref{fig:7}(f) for $f = 6.35$. 
For $7 < f < 8$, we mostly find labyrinth type grain boundary states
such as that illustrated in 
Fig.~\ref{fig:7}(g) for $n = 7.6$, while for
$8 < f < 9$, the system mostly exhibits isolated
defects in a triangular lattice, with
some instances of stripelike domains as shown in 
Fig.~\ref{fig:7}(h) for $f = 8.8$.  

In our study we have only considered square pinning arrays; however, 
the nature of the stripe and grain boundary
states may change considerably for the case of 
hexagonal pinning arrays. 
Additionally, we have only explored the static states, but it is
likely that the effective friction experienced by the particles under
an applied drive can strongly affect the ordering of the particles,
as has been recently demonstrated for domain wall formation
in systems of colloidal particles 
interacting with periodic pinning arrays near the $f=1$ filling \cite{22,20}.   

\section{Conclusions}

We have shown that a rich variety of distinct defect patterns 
occur for colloidal particles interacting
with a square muffin tin potential array. 
The system forms a hexagonal lattice for four colloidal particles per
pinning site and a square lattice for five colloidal particles per
pinning site, and for intermediate fillings we identify
a domain wall regime,
a crystalline stripe regime, and a disordered patch regime. 
We show how these different regimes can be
be distinguished with Voronoi diagrams, grain orientation 
measurements, and energy dispersion. 
These ordered defect patterns arise in the muffin tin potential due to the
fact that
a fraction of the colloidal particles are located
in the interstitial regions between the pinning sites 
and have more mobility than particles on an egg-carton potential.
We show that the pattern formation of the grain boundaries also occurs for
a range of other fillings up to 
nine colloidal particles per trap, with the 
most prominent stripe crystalline states forming when there are between
four and five colloidal particles per trap.
Our results may provide a new method for tailoring defect structures in 
colloidal particle systems.       

\section{Acknowledgements}
This work was carried out under the auspices of the 
NNSA of the 
U.S. DoE
at 
LANL
under Contract No.
DE-AC52-06NA25396.
D.M. and J.A. received support from the ASC Summer Workshop program at LANL.

%The \balance command can be used to balance the columns on the final page if desired. It should be placed anywhere within the first column of the last page.

%\balance

\footnotesize{

}

\end{document}